\documentclass[%
10pt,								% Schriftgröße (12pt, 11pt (Standard))
bibtotocnumbered           %Literaturverzeichnis nummeriert
]
{scrartcl}

\usepackage[ngerman]{babel}
\usepackage[T1]{fontenc}
\usepackage[ansinew]{inputenc}
\usepackage{amssymb}
\usepackage{amsmath}
\usepackage{amsthm}
\usepackage{a4wide}  %Textgröße
\usepackage{longtable}

\usepackage{lmodern} %Type1-Schriftart für nicht-englische Texte

\usepackage{graphicx} %%Zum Laden von Grafiken
\begin{document}

\parindent0pt     %% keine Einrückungen am Absatzanfang
\parskip.7ex      %% kleiner Abstand zwischen Absätzen

%%%%%%%%%%%%%%%%%%%%%%%%%%%%%%%%%%%%%%%%%%%%%%%%%%%%%%%%%%%%%%%%%%%%%%%
%% Ihr Artikel                                                       %%
%%%%%%%%%%%%%%%%%%%%%%%%%%%%%%%%%%%%%%%%%%%%%%%%%%%%%%%%%%%%%%%%%%%%%%%

%% eigene Titelseitengestaltung %%%%%%%%%%%%%%%%%%%%%%%%%%%%%%%%%%%%%%%    
%\begin{titlepage}
%Einsetzen der TXC Vorlage "Deckblatt" möglich
%\end{titlepage}

%% Angaben zur Standardformatierung des Titels %%%%%%%%%%%%%%%%%%%%%%%%
%\titlehead{Titelkopf }
%\subject{Typisierung}
\title{Zwei Anwendungen des Paillier-Kryptosystems: Blinde Signatur und Three-Pass-Protocol}
\author{Anselme Tueno\footnote{anselme.tueno@googlemail.com}}
%\and{Der Name des Co-Autoren}
%\thanks{Fußnote}			% entspr. \footnote im Fließtext
\date{13. September 2009}							% falls anderes, als das aktuelle gewünscht
%\publishers{Herausgeber}

%% Widmungsseite %%%%%%%%%%%%%%%%%%%%%%%%%%%%%%%%%%%%%%%%%%%%%%%%%%%%%%
%\dedication{Widmung}

\maketitle 						% Titel wird erzeugt
%\thispagestyle{empty} %%Keine Kopf-/Fusszeilen auf der ersten Seite (herausgenommen MP)

%% Zusammenfassung nach Titel, vor Inhaltsverzeichnis %%%%%%%%%%%%%%%%%
\begin{abstract}%\textbf{Abstract.}
\noindent Basierend auf dem Kryptosystem von Paillier \cite{Pail99} und dem damit eingeführten Problem
der zusammengesetzten Residuenklasse werden in diesem Artikel zwei
krypto\-graphische Verfahren vorgeschlagen. Zunächst wird die Signatur von
Paillier in ein blindes Signatur\-verfahren umgewandelt. Dabei wird aus einer
verblendeten Nachricht eine blinde Sig"-natur erzeugt, so dass man dann in der
Lage ist, daraus eine gültige Signatur der ursprünglichen Nachricht zu
berechnen. Des Weiteren wird mit der homomorphen Eigen\-schaft des Kryptosystems
von Paillier ein sogenanntes Three-Pass-Protocol -- auch No-Key-Protocol genannt
-- entwickelt. Das Protokoll erlaubt zwei Teilnehmern, eine Nachricht ohne
vorherigen Schlüsselaustausch vertraulich zu übermitteln. 

\end{abstract}

%% Erzeugung von Verzeichnissen %%%%%%%%%%%%%%%%%%%%%%%%%%%%%%%%%%%%%%%
%%%\tableofcontents			% Inhaltsverzeichnis
%\listoftables				% Tabellenverzeichnis
%\listoffigures				% Abbildungsverzeichnis

%% Der Text %%%%%%%%%%%%%%%%%%%%%%%%%%%%%%%%%%%%%%%%%%%%%%%%%%%%%%%%%%%
\section{Grundlagen}
Um den Artikel verständlicher zu machen, werden in diesem Abschnitt relevante
mathematische Grundlagen des Kryptosystems von Paillier eingeführt. Alle
Berechnungen werden modulo $n^2$ durchgeführt, die Berechnungsgruppe ist also
$\mathbb{Z}^{*}_{n^2}$ mit Eins\-element. 

\medskip
\textbf{Definition 1.1.} Ein Zahl $z \in \mathbb{Z}^{*}_{n^2}$ ist ein
\textit{$n$-Residuum} modulo $n^2$, wenn es eine Zahl $y \in
\mathbb{Z}^{*}_{n^2}$ gibt, so dass gilt: 

$$
z = y^n \bmod  n^2. \label{eqn10}
$$
Da der \textit{Residuumsgrad} die zusammengesetzte Zahl $n$ ist, spricht man
von \textit{zusammengesetztem Residuu}m. Die Menge der $n$-Residuen ist eine
multiplikative Untergruppe von $\mathbb{Z}^{*}_{n^2}$ der Ordnung $\phi(n)$.
Die Zahl $y$ heißt \textit{Residuumswurzel} oder kurz \textit{Wurzel}. Jedes
$n$-Residuum $z$ hat genau $n$ Residuumswurzeln. Die Menge aller Wurzel von $z$
wird mit Roots($z$) bezeichnet. Es gibt genau eine Wurzel in Roots($z$), die
kleiner als $n$ ist und mit $\sqrt[n]{z} = z^{1/n \bmod \lambda} \bmod n$
bezeichnet wird: die \textit{Hauptwurzel}. Die Wurzel des Einselememts sind
Zahlen der Form $(1+n)^x\ =\ 1 + xn\bmod n^2$. 

\bigskip\noindent
\textbf{Lemma 1.2.} Es gilt Roots(1) = $\left\langle 1 + n \right\rangle$.

\begin{proof}[Beweis]
Roots(1) ist offensichtlich eine Untergruppe von $\mathbb{Z}^{*}_{n^2}$ und es
gilt $ord(1 + n) = n$. Nun sei $g \in \left\langle 1 + n \right\rangle$, dann
gilt $g = (1+n)^k$ und $ord(g) = \dfrac{n}{ggT(n, k)}$, die ein Teiler von $n$
ist, d.h. $g \in$ Roots(1). Ferner haben die beiden Menge die gleiche Ordnung,
woraus die Behauptung folgt.
\end{proof}

Nun sei $u = (1 + n)^t \in \mathrm{Roots}(1)$. Da $t \leq n$ gilt, ist $t$ der diskrete
Logarithmus von $u$ zur Basis $1 + n$. Aus der Gleichung $u = (1 + n)^t = 1 +
tn\bmod n^2$ folgt $t = \dfrac{u - 1}{n}$. Daher ist für alle $u \in \mathrm{Roots}(1)$
die Funktion $$L: u \mapsto \frac{u - 1}{n}$$ wohldefiniert.

\bigskip\noindent
\textbf{Definition 1.3.} Eine \textit{Residuenbasis} ist eine Zahl $g \in
\mathbb{Z}^{*}_{n^2}$, deren Ordnung durch $n$ teilbar ist.

Diese Definition reicht aber nicht aus, um eine Residuenbasis algorithmisch
auswählen zu können. Folgende Lemmata sollen deshalb eine effiziente Auswahl
einer Residuenbasis ermöglichen. 

\bigskip
\textbf{Lemma 1.4.} Sei $g = (1 + n)^k$. Dann ist $g$ genau dann eine Residuenbasis wenn gilt: $ggT(k, n) = 1$.

\begin{proof}[Beweis]
Ist $g = (1 + n)^k$ eine Residuenbasis, dann ist $n$ ein Teiler von $ord(g)$.
Gelte nicht $ggT(k, n) = 1$, dann ist $k$ entweder gleich $p$ oder $q$ oder $n$.

Falls $k = p$ (bzw. $q$ ) dann gilt $ord(g) = q$ (bzw. $p$). Falls $k = n$,
dann gilt $ord(g) = 1$. Dies steht im Widerspruch zur Annahme und somit muss $ggT(k,
n) = 1$ gelten.

Gilt umgekehrt $ggT(k, n) = 1$, dann gilt
$$
ord((1 + n)^k)\ =\ \frac{ord(1 + n)}{ggT(ord(1 + n), k)}\ =\ \frac{n}{ggT(n, k)} = n.
$$
\\
\end{proof}

\textbf{Lemma 1.5.} Eine Zahl $g \in \mathbb{Z}^{*}_{n^2}$ ist genau dann eine Residuenbasis wenn gilt: 

\begin{eqnarray}
ggT(L(g^{\lambda}\bmod n^{2}),\ n) = 1.
\end{eqnarray}
\begin{proof}[Beweis]
Nach Carmichael (Lemma 1.13) gilt $g^{\lambda}\ \equiv\ 1\bmod n$. Daraus folgt
$g^{\lambda}\ \equiv\ 1 + kn\bmod n^2$ für ein $k \in \mathbb{Z}_n$.
$L(g^{\lambda}\bmod n^{2})$ liefert $k$ und das voherige Lemma beendet den Beweis.
\end{proof}

Residuenbasen sind der Grundbaustein für die Verschlüsselungsfunktion des
Kryptosystems von Paillier. Sei $g \in \mathbb{Z}^{*}_{n^2}$ eine
Residuenbasis, dann kann man zeigen , dass sich jedes Element $w \in
\mathbb{Z}^{*}_{n^2}$ als $w = g^x \cdot y^n$ eindeutig darstellen lässt \cite{Pail01}. Dies
wird als \textit{Repräsentation} von $w$ nach $g$ bezeichnet. Folgende Funktion
ist von daher wohldefiniert:
\begin{eqnarray*}
E_g : \mathbb{Z}_n \times \mathbb{Z}^{*}_{n} &\mapsto& \mathbb{Z}^{*}_{n^2} \\
(x, y) &\mapsto& g^x \cdot y^n\bmod n^2\\
\end{eqnarray*}

\textbf{Lemma 1.6.} Die Funktion $E_g$ ist bijektiv.

\begin{proof}[Beweis]
Es gilt $\left|\mathbb{Z}_n \times \mathbb{Z}^{*}_{n}\right| = n \cdot \phi(n) = \left|\mathbb{Z}^{*}_{n^2}\right|$. Es bleibt also die Injektivität zu zeigen.
Seien $(x_1, z_1), (x_2, z_2) \in \mathbb{Z}_n \times \mathbb{Z}^{*}_{n}$, mit $z_1$ und $z_2$ $n$-Residuen und angenommen $E_g(x_1, z_1) = E_g(x_2, z_2)$. Daraus folgt $g^{x_1} \cdot z_1 = g^{x_2} \cdot z_2\ \Leftrightarrow\ g^{x_2-x_1} \cdot z_2 \cdot z^{-1}_1 = 1$. Aus der Eindeutigkeit der Darstellung des Einselements folgt $x_1 = x_2$ und $z_1 = z_2$.\\                
\end{proof}

Die Funktion $E_g$ ist die Verschlüsselungsfunktion des Kryptosystems von
Pailier. Mit $D_g$ soll ihre Umkehrfunktion bezeichnet werden.

\textbf{Definition 1.7.} Sei $g$ eine Residuenbasis und $w \in
\mathbb{Z}^{*}_{n^2}$. Die $n$-Residuenklasse $ \left[\left[w\right]\right]_g$
von $w$ nach $g$ ist die eindeutige Zahl $x$, so dass gilt: $w = g^x \cdot z$.
Die Zahl $z$ ist das dazugehörige Residuum $res_g w$.

Aus dieser Definition lassen sich folgende Eigenschaften der Residuenklasse heraustellen.

\medskip
\textbf{Lemma 1.8.}
\begin{enumerate}
	\item[$i.$] $\forall \ w \in \mathbb{Z}^{*}_{n^2} \ \left[\left[w\right]\right]_g = 0$ genau dann wenn $w$ ein Residuum ist.
	\item[$ii.$] Die Funktion, die jedem Element von $\mathbb{Z}^{*}_{n^2}$ seine Klasse zuweist, ist eine additiver Homomorphismus. Es gilt also:
	 \begin{eqnarray}
	 \forall w_1, w_2\ \in \mathbb{Z}^{*}_{n^2}\ \left[\left[w_1w_2\right]\right]_g = \left[\left[w_1\right]\right]_g + \left[\left[w_2\right]\right]_g\bmod n. 
	 \end{eqnarray}
	\item[$iii.$](Klassenformel) $\forall g_1, g_2$ Residuenbasen und $\forall \ w \in \mathbb{Z}^{*}_{n^2}$ gilt:
	 \begin{eqnarray}
	  \left[\left[w\right]\right]_{g_2} = \left[\left[w\right]\right]_{g_1} \left[\left[g_1\right]\right]_{g_2}\bmod n
	  \end{eqnarray}
\end{enumerate}

\begin{proof}[Beweis]\ \newline

 \begin{enumerate}	
	\item[$i.$] Aus $\left[\left[w\right]\right]_g = 0$ folgt $w = g^0 z = z$ ist ein Residuum. Sei umgekehrt $w$ ein Residuum, dann gilt $w = g^0 w$, und damit gilt $\left[\left[w\right]\right]_g = 0$. 

   \item[$ii.$] Seien $w_1 = g^{c_1} z_1$ und $w_2 = g^{c_2} z_2$. Es gilt: $w_1w_2 = g^{c_1+c_2} z_1z_2$. Aus der Eindeutigkeit der Repräsentation folgt $\left[\left[w_1\right]\right]_g\left[\left[w_2\right]\right]_g = \left[\left[w_1\right]\right]_g + \left[\left[w_2\right]\right]_g \bmod n$.

   \item[$iii.$]Seien $g_1$, $g_2$ Residuenbasen  und $w \in \mathbb{Z}^{*}_{n^2}$. Es gilt
   $w = g_{1}^{c_1} z_1$, $g_1 = g_{2}^{\alpha}\beta$ und \\
   $\alpha c_1 = (\alpha c_1\ div\ n) n + (\alpha c_1\bmod n)$. Daraus folgt\\
   $w = (g_{2}^{\alpha} \beta)^{c_1} z_1 = g_{2}^{\alpha c_1}  \beta^{c_1} z_1 = g_{2}^{(\alpha c_1\ div\ n) n + (\alpha c_1\bmod n)}  \beta^{c_1} z_1 = g_{2}^{\alpha c_1\bmod n}  (g_{2}^{\alpha c_1\ div\ n} )^n \beta^{c_1} z_1$ und dies ist eine Repräsentation von $w$ nach $g_2$, denn $(g_{2}^{\alpha c_1\ div\ n} )^n \beta c_1 z_1$ ist ein Residuum und $\left[\left[w\right]\right]_{g_1}= c_1$ und $\left[\left[g_1\right]\right]_{g_2} = \alpha$. 

\end{enumerate}

\end{proof}

Als Verschlüsselungsfunktion muss $E_g$ die Einwegeigenschaft erfüllen, d.h.
einfach zu berechnen und schwierig zu invertieren sein; die Invertierung soll
aber mit dem geheimen Schlüssel einfach berechenbar sein. Diese Funktion ist
offensichtlich mit der schnellen Potenzierung einfach zu berechnen. Um die
Invertierung durchführen zu können, muss man aus der Repräsentation \mbox{$w = g^x
\cdot y^n$} in der Lage sein, die Residuenklasse $x$ zu berechnen. Darüber
hinaus wird in dem Kryptosystem $y$ zufällig ausgewählt. Dies garantiert die
sogenannte \textit{semantische Sicherheit}: der gleiche Klartext wird auf
verschiedene Chiffretexte abgebildet. Um dann zwei Chiffretexte des gleichen
Klartextes unterscheiden zu können, muss der Angreifer die Residuosität
entscheiden können.

\medskip
\textbf{Definition 1.9. Problem der Residuenklasse}.\ Sei  $w \in \mathbb{Z}^{*}_{n^2}$, berechne  die  Klasse  von  $w$.  Dieses  Problem  wird  mit  CLASS[$n$] bezeichnet.

\medskip
\textbf{Definition 1.10. Problem der Entscheidbarkeit der Residuosität}.\ Sei
$w \in \mathbb{Z}^{*}_{n^2}$, entscheide, ob $w$ ein Residuum ist. Dieses
Problem wird CR[$n$] bezeichnet.

\medskip
Auf diesen beiden Problemen und dem Faktorisierungproblem des RSA-Moduls $n$
basiert die Sicherheit des Kryptosystems von Paillier. Es ist nicht leicht zu
beweisen, das deren Berechnungen schwierig sind; man kann es nur vermuten.

\medskip
\textbf{Vermutung 1.11. Composite Residuosity Assumption}. Wenn die
Faktorisierung von $n$ hart  ist, gibt  es  keinen  Algorithmus, der  CR[$n$]
in polynomialer Zeit löst. Diese Vermutung wird mit CRA bezeichnet und
garantiert die semantische Sicherheit.

\medskip
\textbf{Vermutung 1.12. Computational Composite Residuosity Assumption}. Wenn
die Faktorisierung  von  $n$  hart ist, dann gibt es keinen Algorithmus, der
der CLASS[$n$] in polynomialer Zeit löst. Diese Vermutung wird als CCRA
bezeichnet. Die Komplexität des Problems der Residuenklasse CLASS[$n$]
garantiert die Einwegeigenschaft.

\medskip
Diese Vermutungen garantieren die semantische Sicherheit und die
Einwegeigenschaft. Kennt man aber den geheimen Schlüssel $\lambda$, ist man
in der Lage, die Entschlüsselung einfach durchzuführen.

\medskip

\textbf{Lemma 1.13 Carmichael.} Sei $w \in \mathbb{Z}^{*}_{n^2}$, Es gelten: 

$$w^{\lambda} \ \equiv\ 1 \bmod n \ \mathrm{und} \ w^{n \lambda} \ \equiv\ 1 \bmod n^2.
$$

\medskip

\textbf{Lemma 1.14.} \ Sei $w \in \mathbb{Z}^{*}_{n^2}$, es gilt:

$$
L(w^{\lambda}\bmod n^{2}) = \lambda \left[\left[w\right]\right]_{1 + n}  \bmod n.
$$

\begin{proof}[Beweis]
Aus Lemma 2.4 folgt, dass $1 + n$ eine Residuenbasis ist. Das Element $w$ ist also nach $1 + n$ repräsentierbar und es gibt eine eindeutiges Paar $(a, y) \in \mathbb{Z}_n \times \mathbb{Z}^{*}_{n}$, so dass 
$w = (1 + n)^a y^n \bmod n^2$ gilt. Also $\left[\left[w\right]\right]_{1 + n} = a$.\\
Nun  
\begin{displaymath}
 \begin{array}{ccc}
w^{\lambda} & = & (1 + n)^{a \lambda} y^{n \lambda}  \\
 & = & (1 + n)^{a \lambda} y^{n \lambda} \\
 & = & 1 + a \lambda n \bmod n^2
\end{array} 
\end{displaymath}
Daraus folgt:
$$
L(w^\lambda \bmod n^2) = \frac{1 + a \lambda n - 1}{n} \bmod n  =  a \lambda \bmod n = \lambda \left[\left[w\right]\right]_{1 + n} \bmod n
$$

\end{proof}

\medskip

\textbf{Korollar 1.15.} \ Seien $g$ Residuenbasis und $w \in \mathbb{Z}^{*}_{n^2}$, Es gilt:

\begin{eqnarray}
\left[\left[w\right]\right]_{g} = \frac{L(w^{\lambda} \bmod n^2)}{L(g^{\lambda} \bmod n^2)} \bmod n.
\end{eqnarray}
\begin{proof}[Beweis]
Wegen der Klassenformel von Lemma 1.8 (3) gilt $1 = \left[\left[1+n\right]\right]_{1 + n} = \left[\left[w\right]\right]_g \left[\left[g\right]\right]_{1 + n}$. Daraus folgt, dass $\left[\left[g\right]\right]_{1 + n} = \left[\left[1+n\right]\right]_{g}^{-1} \bmod n$ invertierbar $mod \ n$ ist. Damit ist auch $L(g^{\lambda} \bmod n^2) = \lambda \left[\left[g\right]\right]_{1 + n}$ invertierbar, da $ggT(\lambda, n) = 1$ gilt.
Aus der Klassenformel wiederum gilt $\left[\left[w\right]\right]_{1 + n} = \left[\left[w\right]\right]_g \left[\left[g\right]\right]_{1 + n}$. Daraus folgt:
$$
\left[\left[w\right]\right]_g =  \frac{\left[\left[w\right]\right]_{1+n}}{\left[\left[g\right]\right]_{1+n}} =\frac{\lambda \left[\left[w\right]\right]_{1+n}}{\lambda \left[\left[g\right]\right]_{1+n}} = \frac{L(w^{\lambda} \bmod n^2)}{L(g^{\lambda} \bmod n^2)}.  
$$                                                                              

\end{proof}
Der letzte Satz gibt also eine Formel (4) für die Berechnung der Klasse eines
Elements $w \in \mathbb{Z}^{*}_{n^2}$, die benutzt wird, um das Verfahren zu
entschlüsseln. Die Berechnung ist offensichtlich bei bekanntem geheimem
Schlüssel $\lambda$ leicht durchzuführen.

\section{Probabilistisches Schema}
Im folgendem Kryptosystem wird eine Residuenbasis mit der Gleichung (1)
ausgewählt, eine Nachricht mittels der Funktion $E_g$ verschlüsselt und der
Formel aus Gleichung (4) entschlüsselt. Auf Basis dieses Kryptosystems hat
Paillier auch eine Einweg-Falltürpermutation und ein Signaturverfahren
entwickelt. 

\begin{center}
\begin{tabular}{|rl|}
\hline
\rule{0mm}{2.7ex}\textbf{Schlüsselgenerierung} & Große Primzahlen $p$ und $q$ selber Länge $(p \neq q): n = pq$,~~~ \\
                     & Residuenbasis $g \in_R \mathbb{Z}^{*}_{n^2}$ mit $ggT(L(g^{\lambda} \bmod n^2), n) = 1$ \\
                     & Öffentlicher Schlüssel: $(n, g)$, \\
\rule[-2mm]{0mm}{1ex}                     & Privater Schlüssel: $(p, q)$ bzw. $\lambda = kgV(p - 1, q - 1)$ \\
\hline%\hline
\rule{0mm}{2.7ex}\textbf{Verschlüsselung}      & Klartext: $m \in \mathbb{Z}_n$, \\
                     & Zufallszahl: $x \in_R \mathbb{Z}_{n}^{*}$, \\
\rule[-2mm]{0mm}{1ex}                     & Chiffretext: $c = g^m x^n \bmod n^2$ \\
\hline%\hline
\rule{0mm}{2.7ex}\textbf{Entschlüsselung}      & Chiffretext: $c < n^2$, \\
                     & Klartext: $m = \dfrac{L(c^{\lambda} \bmod n^2)}{\rule[-2mm]{0mm}{1ex}L(g^{\lambda} \bmod n^2)} \bmod n$\\
\hline
\end{tabular}
\end{center}

\textbf{Bemerkung 2.1.} Das obige Kryptosystem ist wegen der Zufallzahl $x$
probabilistisch. Die Korrektheit wurde mit Korollar 1.15 bewiesen. Die
Sicherheit basiert auf den obigen vorgestellten Problemen.

\medskip
\textbf{Theorem 2.2.} Das probalistische Kryptosystem von Paillier hat genau
dann die Einwegeigenschaft wenn CCRA gilt.

\medskip
\textbf{Theorem 2.3.} Das probalistische Kryptosystem von Paillier ist genau
dann semantisch sicher, wenn CRA gilt.

\begin{proof}[Beweis]
Seien $m_0$, $m_1$ zwei Klartexte und $c$ ein Chiffretext eines von den Beiden.
Um die semantische Sicherheit zu brechen, muss der Angreifer entscheiden
können, wessen Chiffretext $c$ ist. Für $i \in \left\{0, 1\right\}$ ist $c$ ein
Chiffretext von $m_i$ genau dann, wenn $cg^{-m_i} \bmod n_2$ ein $n$-Residuum
ist. Der Angreifer muss also in der Lage sein CRA zu lösen.  
\end{proof}

Da der Klartext $m \in \mathbb{Z}_n$ und $c < n^2$ hat dieses Schema einen
Expansionsfaktor\footnote{Verhältnis zwischen der Länge des Klartextes und der
des Chiffretextes.} von \\
$(log(n))/(log(n^2))  =  1/2$, d.h. der Chiffretext ist doppelt so lang wie
der Klartext. Dies kann verbessert werden indem $m$ aus $\mathbb{Z}_{n^2}$
ausgewählt und als Paar $(m_1, m_2)$ von Zahlen dargestellt wird, wobei
$ggT(m_2, n) = 1$ und $m_1 < n$ gelten müssen. Das resultierende System ist
eine \textit{Einweg-Falltürpermutation}. Paillier hat die $n$-adische
Darstellung von $m$  vorgeschlagen. Es wird zunächst das System wie in
\cite{Pail99} vorgestellt und dann dessen Korrektheit diskutiert.

\begin{center}
\begin{tabular}{|rl|}
\hline
\rule{0mm}{3ex}\textbf{Schlüsselgenerierung} & Große Primzahlen $p$ und $q$ selber Länge $(p \neq q): n = pq$,~~~ \\
                     & Residuenbasis $g \in_R \mathbb{Z}^{*}_{n^2}$ mit $ggT(L(g^{\lambda} \bmod n^2), n) = 1$ \\
                     & Öffentlicher Schlüssel: $(n, g)$, \\
\rule[-2mm]{0mm}{1ex}                      & Privater Schlüssel: $(p, q)$ bzw. $\lambda = kgV(p - 1, q - 1)$ \\
\hline%\hline
\rule{0mm}{3ex}\textbf{Verschlüsselung}      & Klartext: $m \in \mathbb{Z}_{n^2}$, $m_1 = m \bmod n$ und $m_2 = m \ div \ n$, \\
\rule[-2mm]{0mm}{1ex}                      & Chiffretext: $c = g^{m_1} m_{2}^{n} \bmod n^2$ \\
\hline%\hline
\rule{0mm}{3.5ex}\textbf{Entschlüsselung}      & Schritt1: $m_1 = \frac{L(c^{\lambda} \bmod n^2)}{L(g^{\lambda} \bmod n^2)} \bmod n$\\
	& Schritt 2: $z = cg^{-m_1} \bmod n$, \\
	&	Schritt 3: $m_2 = z^{1/n \bmod \lambda} mod \ n$, \\
\rule[-2mm]{0mm}{1ex}	&	Schritt 4: Klartext $m = m_2 n + m_1$ \\
\hline
\end{tabular}
\end{center}

\medskip
Die Entschlüsselung der Einweg-Falltürpermutation wird in vier Schritten
durchgeführt. Im ersten Schritt wird $m_1$ mit (4) berechnet. Das Residuum $z =
res_g c$ wird im zweiten Schritt berechnet. Im dritten Schritt wird $m_2$ als
die Hauptwurzel von $z$ berechnet. Die $n$-adische Entwicklung von $m$ wird im
vierten Schritt berechnet und damit ist der Klartext gewonnen. Mit der
$n$-adischen Darstellung $m = m_{2}n + m_1$ gilt aber nicht immer $ggT(m_2, n) = 1$
und damit kann das Schema nicht für alle Klartexte korrekt sein.

\medskip
\textbf{Bemerkung 2.4.} Ist der Klartext $m$ kleiner $n$, dann gilt $m_2 = 0$
und damit $c = 0$. Daher können Klartexte kleiner als $n$ nicht verschlüsselt
werden. Das Schema ist in diesem Fall deterministisch und kann daher nicht
semantisch sicher sein.

\section{Eigenschaften}
In diesem Abschnitt werden zwei Eingenschften des probalistischen Kryptosystems
von Paillier vorgestellt, die wichtig sind, um kryptographische Protokolle zu
entwickeln.

\subsection{Homomorphie}
Bezüglich  der Anwendungen ist diese Eigenschaft die wichtigste und folgt aus  dem  additiven Homomorphismus von Lemma 1.8 (2).

Seien $m, m_1, m_2 \in \mathbb{Z}_n, k \in \mathbb{N}$ und $g$ eine Residuenbasis, so gelten folgende Gleichungen:
   
\begin{eqnarray}
D_g (E_g(m_1)E_g(m_2) \bmod n^2) & = & m_1 + m_2 \bmod n  \\
D_g (E_g (m)^k \bmod n^2) & = & km \bmod n  \\
D_g (E_g (m_1)g^{m_2} \bmod n^2) & = & m_1 + m_2 \bmod n \\
D_g (E_g (m_1)^{m_2} \bmod n^2) & = & m_1 m_2 \bmod n 
\end{eqnarray}

Diese Eigenschaft ist notwendig, um kryptographische Protokolle zu entwickeln,
bei denen mit verschlüsselten Daten gerechnet werden muss. Dazu  zählen
beispielsweise elektronische Wahlen, Secret Sharing, Copyright etc. Das im  5.
Abschnitt  vorgestellte Three-Pass-Protokoll basiert auf Gleichung (8).

\subsection{Self-Blinding}
Mit dieser Eigenschaft ist es möglich, einen Chiffretext in einen anderen
Chiffretext umzuwandeln, ohne den jeweiligen Klartext zu kennen.

Für alle $m \in \mathbb{Z}_n, \ x \in \mathbb{Z}_{n}^{*}, \ r \in \mathbb{N}$ und $g$ eine Residuenbasis gelten die Beziehungen:
\begin{eqnarray}
D_g (E_g (m) x^n \bmod n^2)  & = & m\\
D_g (E_g (m) g^{nr} \bmod n^2) & = & m
\end{eqnarray}
Wie blinde Signaturen ermöglicht diese Eigenschaft die Entwicklung von
kryptographischen Protokollen wie elektronischen Zahlungssystemen, bei denen
die Anonymität gefordert wird.

\section{Digitale Signatur}
Um mit dem probabilistischen Kryptosystem signieren zu können, konstruiert
Paillier folgendes Signaturverfahren, bei dem die Schlüsselgenerierung erhalten
bleibt.

\begin{center}
\begin{tabular}{|rl|}
\hline
\rule{0mm}{2.7ex}Signaturalgorithmus & Klartext: $m \in \mathbb{Z}_{n^2}^{*}$, \\
                    & $s_1 = \dfrac{L(m^{\lambda} \bmod n^2)}{L(g^{\lambda}\bmod n^2)}  \bmod n$, \\
			  	          & $s_2 = (mg^{-s_1})^{1/n \bmod \lambda} \bmod n$,\\
 \rule[-2mm]{0mm}{1ex}                   & Signatur: $\sigma(m) = (s_1, s_2)$ \\
\hline%\hline
\rule[-2mm]{0ex}{0ex}\rule{0mm}{2.7ex}Verifikationsalgorithmus &  $m \ ^{?}_{=} \ g^{s_1} s_{2}^{n} \bmod n^2$\\
\hline%\hline
\rule[-2mm]{0ex}{0ex}\rule{0mm}{2.7ex}Korrektheit & $g^{s_1} s_{2}^{n}  = g^{s_1} m g^{-s_1} = g^{s_1} g^{-s_1} m  = m$~~~\\
\hline
\end{tabular} 
\end{center}

\medskip
In der Praxis wird nicht der Klartext $m$ selbst signiert, sondern dessen
Hashwert. Dafür wird eine Hashfunktion $h: \mathbb{N} \mapsto \ \left\{0,1\right\}^k \subset \ \mathbb{Z}_{n^2}^{*}$
benötigt. Man ersetze also im obigen Schema $m$ durch $h(m)$.

Manchmal ist es wünschenswert, dass der Inhalt der Nachricht dem Signierer
geheim bleibt. Dieses Problem wurde in \cite{Chau82} von David Chaum mit blinden
digitalen Signaturen gelöst. Dabei signiert ein Teilnehmer B eine Nachricht für
einen anderen Teilnehmer, ohne den Inhalt der Nachricht nachvollziehen zu
können.

\medskip
\textbf{Definition 4.1.} Ein blindes digitales Signaturschema ist ein Signaturschema mit folgenden zusätzlichen Eigenschaften:

\begin{itemize}
	\item Es gibt zwei Teilnehmer: den Signierer und den Provider
	\item Nur der Signierer kann eine Signatur erzeugen
	\item Der Provider kennt allein eine invertierbare Funktion, mit der er seine Nachricht blenden kann, bevor sie zum Signieren geschickt wird.
\end{itemize}

Sei $n$ ein RSA-Modul, $e$ bzw. $d$ der öffentliche bzw. private RSA-Schlüssel.
Folgendes Beispiel von D. Chaum erläutert, wie man mit RSA eine Nachricht $m$
blind signieren kann.

\medskip
\textbf{Beispiel 4.2.} Der Provider will seine Nachricht $m$ blind signieren. Er blendet $m$, indem er eine Zufallzahl $r \in_R \mathbb{Z}_{n}^{*}$ auswählt und $M = m r^e \bmod n$ berechnet. Der Signierer bekommt die Nachricht $M$ und versieht sie mit der RSA-Signatur $\sigma(M)$. Um die Signatur der ursprünglichen Nachricht $m$ zu bekommen,  multipliziert der Provider $\sigma(M)$  mit $r^{-1}$. Denn es gilt:

$$  \sigma(M) = M^d = (m r^e)^d = m^{d}r^{ed} = m^{d}r = \sigma(m)r \bmod n.$$
Es kann ohne zusätzliche Schwierigkeit verifiziert werden, dass $\sigma(m)$
tatsächlich eine Signatur von $m$ ist. Das Prinzip von Chaum wird nun auf das
Signaturschema von Paillier angewendet und führt damit ein Signaturverfahren
ein, das ermöglicht, blinde Signaturen zu erzeugen.

\medskip
\textbf{Theorem 4.3.} Seien $x \in_R \mathbb{Z}_{n}^{*}$, $m \in_R \mathbb{Z}_{n^2}^{*}$ eine Nachricht und $M = mx^{n}$ eine Verblendung von $m$. Ferner sei $\sigma(M) = (s_1, s_2)$ die Paillier-Signatur von $M$. Dann ist $\sigma(m) = (s_1, s_2 x^{-1} \bmod n)$ eine gültige Signatur von $m$.

\begin{proof}[Beweis]
Es gilt: $$M^{\lambda} \bmod n^2 = (mx^n )^{\lambda} \bmod n^2 = m^{\lambda}x^{n\lambda}\bmod n^2 = m^{\lambda}\bmod n^2,$$ da aus dem Satz von Carmichael $x^{n\lambda}\ \equiv \ 1 \bmod n^2$ gilt.

Also die Signatur $\sigma(M) = (s_1, s_2)$ mit:
$$s_1 = \frac {L(M^{\lambda} \bmod n^2)}{L(g^{\lambda} \bmod n^2)} \bmod n = \frac {L(m^{\lambda} \bmod n^2)}{L(g^{\lambda} \bmod n^2)} \bmod n$$ und

\begin{displaymath}
 \begin{array}{ccc}
s_2 & = & (mx^{n}g^{-s_1})^{1/n \bmod {\lambda}} \bmod n  \\
 & = & (mg^{-s_1})^{1/n \bmod {\lambda}} (x^n)^{1/n \bmod {\lambda}} \bmod n \\
 & = & x(mg^{-s_1})^{1/n \bmod \lambda}\bmod n
\end{array} 
\end{displaymath}

\medskip
Eine gültige Signatur von $m$ ist also $\sigma(m) = (s_1, s_2 x^{-1} \bmod n)$.
\end{proof}

\section{Three-Pass-Protocol}
\subsection{Definition}
Ein Three-Pass-Protocol ist ein Protokoll, das erlaubt eine Nachricht
vertraulich ohne Schlüssel\-austausch zu senden. Sender und Empfänger müssen
dabei genau drei verschlüsselte Nachrichten austauschen, daher der Name. Das
Protokoll benutzt eine Chiffrierfunktion $E$ mit privatem Chiffrierschlüssel
$e$ und eine Dechiffrierfunktion $D$ mit ebenfalls privatem
Dechiffrierschlüssel $d$, so dass $D(d, E(e, m)) = m$ gilt. Die Verschlüsselung
soll kommutativ sein, das heißt es soll für alle Schlüssel $a$ und $b$ und alle
Nachrichten $m$ gelten: $E(a, E(b, m)) = E(b, E(a, m))$ und umgekehrt
$D(d,E(k,E(e,m))) = D(d,E(e,E(k,m))) = E(k,m)$. 

Ursprünglich wurde das Protokoll von Adi Shamir eingeführt und wird als Shamir
No-Key-Protocol bezeichnet, weil Sender und Empfänger keinen Schlüssel
austauschen. Es wird zunächst das Protokoll von Shamir vorgestellt und im
Anschluss daran ein No-Key-Protocol basierend auf dem Kryptosystem von Paillier
eingeführt.

\subsection{Three-Pass-Protocol von Shamir}
Jeder Teilnehmer besitzt zwei private Schlüssel jeweils für Ver- und
Entschlüsselung. Das Protokoll basiert auf Potenzierung modulo einer großen
Primzahl $p$. Teilnehmer $T$ erzeugt für die Kommunikation einen Schlüssel
$e_T$ mit $e_T < p - 1$ welcher relativ prim zu $p - 1$ ist. Dann bestimmt er
das Inverse $d_T$ von $e_T$ modulo $p - 1$, es gilt also $e_T * d_T \equiv \ 1
\bmod p - 1$. Aufgrund des kleinen Satzes von Fertmat gilt für alle
Nachrichten $m$:
$$
(m^{e_T})^{d_T} = m^{e_T*d_T} = m^{k*(p-1)+1} \equiv m \bmod p
$$
Das Protokoll ist dem Verfahren von Diffie und Hellman sehr ähnlich, jedoch
ohne Schlüsselaustausch. $A$ und $B$ seien die zwei Teilnehmer des Verfahrens.
$A$ möchte eine Nachricht $m$ an $B$ senden. Er berechnet
$M_1 \ = \ m^{e_A} \bmod p$ und sendet $M_1$ zu $B$. $B$ potenziert $M_1$ mit seinem privaten Schlüssel und
sendet $M_2 \ = \ M_1^{e_B} = m^{e_A * e_B} \bmod p$ zu $A$. $A$ entschlüsselt $M_2$ indem er
$M_3 = M_2^{d_A} = (m^{e_A * e_B})^{d_A} \bmod p = m^{e_B} \bmod p$ berechnet. Dann erhält $B$ $M_3$ und
kann nun die ursprüngliche Nachricht mit $M_3^{d_B} = (m^{e_B})^{d_B} \bmod p = m \bmod p $ berechnen.

Eine Verbesserung ist das Massey-Omura Kryptosystem mit Potenzierung in dem
Galoiskörper $GF(2^{n})$ . Die Sicherheit ist gewährleistet, da der Angreifer
keine Information über die Nachrichten $M_1$, $M_2$ und $M_3$ herleiten kann.
Beide Verfahren sind anfällig gegen Man-In-The-Middle-Attacken. Aufgrund des
Problems des diskreten Logarithmus ist es jedoch unmöglich, die Schlüssel der
Teilnehmer aus einem der ausgetauschten Chiffretexte zu erschließen.

\subsection{Three-pass-Protocol mit Paillier}
In diesem Abschnitt wird ein Three-Pass-Protocol basierend auf dem
Paillier-Kryptosystem vorgeschlagen. Zum Verfahren von Shamir gibt es zwei
wesentliche Unterschiede:

\begin{itemize}
	\item das hier beschriebene Protokoll benutzt anstelle der Kommutativität die homomorphe
       Eigen\-schaft des Paillier-Kryptosystems. Denn es gilt:
       $$D( (E(m_1))^{m_2} \bmod n^{2}) \ = \ m_1m_2 \bmod n.$$
   \item nur der Sender besitzt Schlüssel, und nur er muss verschlüsseln. Der
      Empfänger berechnet nur eine modulare Potenz und am Ende eine Multiplikation
      mit einer Inversen.
\end{itemize}

$A$ möchte eine Nachricht $m_1$ zu $B$ senden. 

\begin{enumerate}
	\item $A$ verschlüsselt $m_1$ mit Paillier, in dem er $M_1 \ = \ g^{m_1}y^{n} \bmod n^{2}$ berechnet. Er sendet $M_1$ zu $B$.
	\item $B$ wählt eine Nachricht $m_2 < n$ mit ggT($m_2$, $n$) = 1 und damit potenziert er
       $M_1$:
       $$M_2 = M_1^{m_2} \bmod n^{2}  = \ (g^{m_1}y^{n})^{m_2} \bmod n^{2} = \ g^{m_1m_2}(y^{m_2})^{n} \bmod n^{2}$$
       $B$ sendet $M_2$ zurück zu $A$.
	\item Mit der Entschlüsselung von Paillier berechnet $A$ $M_3 = D(M_2) = m_1m_2 \bmod n$.
	\item $B$ bekommt die Zahl $M_3$ und berechnet $M_3*m^{-1}_2 \bmod n$, und es kommt $m_1$ heraus.
\end{enumerate}

\section{Sicherheit des Protokolls}\label{Sicherheit}
Die Nachrichten $M_1$ und $M_2$ sind Paillier-Chiffres und somit unter den
Annahmen des zusammengesetztem Residuums sicher. Aus $M_3$ kann der Angreifer
vermutlich keine Information über $m_1$ bekommen, wenn $m_2$ zufällig
ausgewählt wird. Eine bessere Sicherheit erhält man indem das Prinzip von
David Chaum benutzt wird. Dabei wählt $B$ nicht irgendeine Zahl $m_2$ sondern
er wählt $m_2$ als $n$-Residuum modulo $n$. Das heißt $B$ wählt zufällig
eine Zahl $x \in_R \mathbb{Z}^{*}_n$ und berechnet $m_2 \ = \ x^{n} \bmod n$.
Dies führt dazu, dass $M_3$ die Form $M_3  = \ m_1x^{n} \bmod n$
hat, die nach Chaum eine Verblendung von $m_1$ und somit sicher ist.

Das Protokoll ist wie beim Diffie-Hellman-Schlüsselaustausch und Shamirs No-Key
Protocol anfällig gegen Man-In-The-Middle-Attacken, sobald ein Angreifer in der
Lage ist, Nachrichten zu erzeugen bzw. abzufangen und zu ersetzen. Das Problem
kann durch ein zusätzliches Authentifizierungs-Protokoll behoben werden.

\section{Zusammenfassung}
In dieser Arbeit wurden zwei kryptographische Verfahren eingeführt, die aus dem
homomorphen Kryptosystem von Paillier basieren.

Das Signaturschema  von Paillier wurde mit dem  gleichen Prinzip von David Chaum 
für blinde RSA-Signaturen  in  ein  blindes  Signaturschema umgewandelt. Dabei
blendet man eine Nachricht $m$ mit einem zufällig gewählten Residuum $x^n$,
bevor sie zum Signierer geschickt wird. Wenn man dann die Signatur der
geblendeten Nachricht bekommt, muss nur der zweite Signaturteil mit \mbox{$x^{-1}$ mod $n$}
multipliziert werden, um die Signatur der ursprünglichen Nachricht $m$  zu haben.

Das hier entwickelte No-Key-Protokoll wurde durch die homomorphe Eigenschaft
des Paillier-Verfahrens ermöglicht. Im Gegensatz zu dem No-Key-Protocol von
Shamir muss nur der Sender ver- und entschlüsseln. Der Empfänger wählt zufällig
eine mit $n$ teilerfremde Zahl und berechnet nur eine modulare Potenz
und am Schluss die Multiplikation mit einer modularen Inversen. Das Protokoll
ist anfällig gegen Man-In-The-Middle-Attacken und benötigt daher zusätlich eine
Authentifizierung. Eine andere Lösung bieten Quanten-Algorithmen. In
\cite{KyYs09} und \cite{Yang03} sind Quantum No-Key-Protocols vorgestellt worden,
die gegen Man-In-The-Middle-Attacken resistent sind, weil das Abhören des
Kanals aufgrund quantenmechanischer Gesetze leicht zu bemerken ist.

Des Weiteren wurde festgestellt, dass die Einweg-Falltürpermutation zwar
theoretisch bijektiv ist und den Expansionfaktor verbessert, jedoch auf Kosten
der Korrektheit. Sie kann nämlich nicht für alle Klartexte korrekt sein. Die
Korrektheit ist für eine Nachricht $m \in \mathbb{Z}_{n^2}$ nur dann
gewährleistet, wenn sich $m$ in $(m_1, m_2)$ darstellen lässt, wobei $m_1 < n$
und $ggT(m_2, n) = 1$ gelten müssen.

\section{Danksagung}
Diese Arbeit ist aus der Diplomarbeit des Autors entstanden. Er möchte deshalb
seiner Betreuerin Prof. Dr. Elena Fimmel sehr herzlich bedanken, für den
Vorschlag eines so interessanten Themas, ihr Verständnis, und die Mühe diese
Arbeit zu lesen und zu begutachten.

\section[Notationen]{Notationen}
$kgV(a, b)$: kleinstes gemeinsames Vielfaches von $a$ und $b$\\
$ggT(a, b)$: größter gemeinsamer Teiler von $a$ und $b$ \\
$n\ =\ pq$: RSA-Zahl, wobei $p$ und $q$ zwei ungleiche große Primzahlen sind \\
$\phi(n)$: Eulersche $\phi$-Funktion von $n$ \\
$\lambda(n)\ =\ kgV(p-1, q-1)$: Carmichaelfunktion\footnote{Aus Übersichtlichkeitgründen wird in dieser Arbeit $\lambda(n)$ nur noch mit $\lambda$ bezeichnet, falls nichts anders erwähnt wird.} von $n$ \\
$ord(g) = min(\left\{e\ :\ g^e\bmod n^2\ =\ 1\right\})$: Ordnung von $g$ modulo $n^2$ \\
$ \left[\left[w\right]\right]$: Residuenklasse von $w$

%\section{Referenzen}

\end{document}